\documentclass{article}
\usepackage{spconf,amsmath,graphicx}
\usepackage{amssymb}
\usepackage{xcolor}
\usepackage[utf8]{inputenc}
\usepackage[normalem]{ulem}
\usepackage{sectsty}
\usepackage{subcaption}
\paragraphfont{\normalfont\bfseries}

\usepackage{paralist}
\usepackage{balance}

\title{Head-related transfer function interpolation with A spherical CNN}
\name{Xingyu Chen, Fei Ma, Yile Zhang, Amy Bastine, Prasanga N. Samarasinghe}
\address{Audio and Acoustic Signal Processing Group, Australian National University\\
E-mail: \{xingyu.chen1, fei.ma, yile.zhang, amy.bastine, prasanga.samarasinghe\}@anu.edu.au}
\begin{document}
\maketitle
\begin{abstract}
Head-related transfer functions (HRTFs) are crucial for spatial soundfield reproduction 
in virtual reality applications. However, obtaining personalized, 
high-resolution HRTFs is a time-consuming and costly task. 
Recently, deep learning-based methods showed promise in interpolating high-resolution 
HRTFs from sparse measurements. 
Some of these methods treat HRTF interpolation as an image super-resolution task, 
which neglects spatial acoustic features. 
This paper proposes a spherical convolutional neural network method for HRTF interpolation.
The proposed method realizes the convolution process by decomposing and reconstructing 
HRTF through the Spherical Harmonics (SHs).
The SHs, an orthogonal function set defined on a sphere, allow the convolution layers to effectively 
capture the spatial features of HRTFs which are sampled on a sphere.
Simulation results demonstrate the effectiveness of the proposed method in achieving accurate
interpolation from sparse measurements, outperforming the SH method and learning-based methods.
\end{abstract}
\section{Introduction}
\label{sec:intro}

The growing popularity of virtual reality and augmented reality applications has sparked a research interest in spatial audio rendering methods. 
One of the most popular methods is to reproduce spatial audio binaurally through headphones.
Binaural reproduction relies on the head-related transfer function (HRTF) \cite{li2020measurement} which represents the scattering effect of human anatomy with respect to the direction of sound. 
The dependence of HRTF on the anatomy of the listener (e.g., pinna, head, and torso shape), makes it highly individual \cite{li2020measurement}. Consequently, accurate HRTF measurement over a large number of directions is desirable for authentic spatial audio reproduction.

However, HRTF measurement is a time-consuming and expensive task~\cite{li2020measurement}. 
To alleviate the burden of HRTF measurement, researchers proposed to interpolate dense HRTF 
from sparsely measured HRTF using methods such as bilinear~\cite{xie2013head} and barycentric interpolation method~\cite{hartung1999comparison, poirier2018anaglyph}.
Spatial domain interpolation was proposed to offer the advantage of preserving HRTF spatial features
and ensuring the interpolation results were physically valid. The HRTF is decomposed onto the spatial basis functions such as spherical harmonics~\cite{zotkin2009regularized, ahrens2012hrtf} and principal components~\cite{xie2012recovery,zhang2020modeling}.

Recently, researchers proposed convolutional neural network (CNN) based methods~\cite{wang2021global,jiang2023modeling,hogg2023hrtf} for HRTF interpolation, and 
showed promising results.
However, CNN was primarily designed for processing signals in a 2D plane. 
Thus, to process HRTF which is defined on a sphere, CNN-based methods require complicated 
plane-sphere projection which does not necessarily capture the spatial features of HRTF. 
There exists the need to address the plane-sphere projection while interpolating 
sparsely measured HRTF with a  CNN.

In this paper, we propose a spherical CNN method \cite{cohen2018spherical,esteves2018learning} for HRTF interpolation. 
The spherical CNN method exploits the spherical harmonics (SHs), an orthogonal function set defined on a sphere, to resolve the plane-sphere projection problem. 
We transform the magnitude of sparsely measured HRTF into SH coefficients which are convoluted with kernel functions expressed in the SH domain.
The kernel function captures the spatial features of HRTF. 
Then we convert the convolution result into the magnitude of dense interpolated HRTF.
We demonstrate the HRTF interpolation performance of the proposed spherical CNN method by comparing it with the spherical harmonic method and learning-based methods.
Our code is available at github.com/xingyuaudio/HRTF-SCNN.

\section{problem formulation}
\subsection{HRTF interpolation}
HRTF depends on the anatomy of a subject and the position of a sound source \cite{li2020measurement}.
Let $H^{\text{left}(\text{right})}_{\text{id}}(\Omega, l)$ denote the HRTF for left(right) 
ear of subject $\text{id} \in\{1, \ldots, \mathrm{ID}\}$, source position $\Omega$, 
and frequency bin $l \in\{1, \ldots, L\}$, $\mathrm{ID}$ is the number of subjects, 
and $L$ is the number of frequency bins.  
Source position $\Omega=(r,\theta,\phi)$ is defined by spherical coordinates,
the radius $r$, the elevation $\theta=$ $[-\frac{\pi}{2}, \frac{\pi}{2}]$ and the azimuth 
$\phi=[0, 2{\pi})$.
Hereafter, the superscript for the left(right) ear and the subscript for the subject number 
are omitted for notation simplicity. 

In this paper, we focus on the far-field HRTF magnitude spectra $H_{\text{M}}(\Omega, l)$, 
i.e., the sound source is at a distance 
$r>1.2$ m from the center of subject head~\cite{li2020measurement}, 
\begin{equation}
H_{\text{M}}(\Omega, l)=20 \log _{10}\left(\left|H(\Omega, l)\right|\right),
\end{equation}
which is a logarithmic scale similar to human auditory perception \cite{logscale}.
The far-field HRTF magnitude spectra $H_{\text{M}}(\Omega, l)$ exhibits minimal dependence on $r$ \cite{li2020measurement},
and thus we redefine $\Omega=(\theta,\phi)$. In this regard, we treat $H_\mathrm{M}(\Omega, l)$ as a function of spherical signal $H: \mathbb{S}^2 \rightarrow \mathbb{R}^L$, where $\mathbb{S}^2$ is a two-dimensional manifold parameterized by $(\theta,\phi)$.

The problem of interest is to interpolate spatially dense HRTFs $\left\{H_{\text{M}}(\Omega, l),\Omega \in \Omega^{\text{dense}} 
\right\}_{}$ from spatially sparse HRTFs $\left\{H_{\text{M}}(\Omega, l), \Omega \in \Omega^{\text{sparse}} \right\} $, where $\Omega^{\text{sparse}} \subset \Omega^{\text{dense}}$.


\subsection{Conventional CNN }
HRTF interpolation was considered as an image super-resolution task by some researchers 
who adopt CNN layers \cite{wang2021global,jiang2023modeling,hogg2023hrtf}.
The conventional CNN layer was originally designed for planar images,
\begin{equation}
(I * k)(x)=\sum_{y \in \mathbb{Z}^2} I(y) k(x-y),
\end{equation}
where  $I$ is the input image on the 2D plane $\mathbb{Z}^2$ and 
$k$ is the learnable convolutional kernel~\cite{cohen2018spherical}. 
This convolution exhibits shift equivariance ~\cite{cohen2018spherical}
\begin{equation}
\label{eq:shift_equiv}
[L_tI]*k=L_t[I*k], 
\end{equation}
where $L_t$ is translation operation and  $L_tI$ is a shifted image.
The shift equivariance Eq.~\eqref{eq:shift_equiv} implies that a kernel $k$ 
shares the same weight parameters at different regions of the image $I$. 
The weight sharing effectively reduces the number of model parameters and 
enhances model generalization capacity. 

However, HRTF is not on a plane but on a sphere where the shift equivariance does not hold.
This fact makes a direct application of conventional CNNs unfeasible  
and necessitates complicated projections of HRTF data~\cite{jiang2023modeling,hogg2023hrtf}.
Considering the limitations of existing methods, we propose a spherical CNN method for HRTF interpolation. 
This method captures spherical features, enabling efficient processing of HRTF data which is distributed on a sphere.

\section{methodology}

In this section, we propose a spherical CNN method for HRTF interpolation. 
We first introduce spherical convolution, which extends the idea of CNN to spherical signals, 
and then introduce spectral transformation, which reduces the computational complexity of spherical convolution, and at last describe the model structure.


\subsection{Spherical convolution}
Spherical CNNs generalize CNN to operate on functions defined on 
a sphere by using spherical convolution
\begin{equation}
(f * k)(\Omega)=\int_{g \in \mathbf{S O}(3)} f(g \eta) k\left(g^{-1} \Omega\right) \mathrm{d} g, 
\label{eq:conv}
\end{equation}
where $f$ and $k$ are spherical functions, $k$ is treated as a learnable convolutional kernel, $g$ is the rotational operator in $\mathbf{S O}(3)$, $\eta$ is the north pole \cite{esteves2018learning}.
The spherical convolution, Eq.~\eqref{eq:conv}, exhibits rotational equivariance
\begin{equation}
[gf]*k= g[f*k], 
\end{equation}
which enables a spherical CNN to share the same set of weight parameters within a kernel $k$ across different regions on a sphere. When considering $H_{\text{M}}$, our goal is use $H_{\text{M}}*k$ to effectively capture spatial features.

\subsection{Spectral transformation}
\label{sec:Spectral}
In this section, we present spectral transformation \cite{esteves2018learning}, a practical method that compute 
the spherical convolution Eq.~\eqref{eq:conv} which does not have a closed-form solution. 
Specifically, the spherical convolution Eq.~\eqref{eq:conv} is approximated  by three steps:
\begin{inparaenum}[(i)]
\item Spherical Harmonic Transform (SHT): decompose spherical domain signals into SH coefficients,
\item multiplication of SH coefficients and kernel coefficients, and
\item Inverse Spherical Harmonic Transform (ISHT): converts the multiplication back into the spherical domain.
\end{inparaenum}



Firstly, 
let the HRTF be measured at directions $\{\Omega_p\}_{p=1}^{P}$ and $L$ frequency bins. 
we decompose HRTF magnitude spectra $H_\mathrm{M}(\Omega, l)$ onto the SHs.
\begin{equation}
\mathbf{H}=\mathbf{Y} \mathbf{a},
\end{equation}
where 
$$
{\tiny
\mathbf{H}=\left[\begin{array}{ccc}
H_{\text{M}}(\Omega_1, 1) & \cdots & H_{\text{M}}(\Omega_1, L)\\
\vdots & \ddots & \vdots \\
H_{\text{M}}(\Omega_P, 1) & \cdots & H_{\text{M}}(\Omega_P, L)
\end{array}\right]}
$$
is a $P\times{(N_H+1)^2}$ matrix containing the real SH basis functions,
$$
{\tiny
\mathbf{Y} =
\left[\begin{array}{ccc}
Y_{00}(\Omega_1) & \cdots & Y_{N_H N_H}(\Omega_1)\\
\vdots & \ddots & \vdots \\
Y_{00}(\Omega_P) & \cdots & Y_{N_H N_H}(\Omega_P)
\end{array}\right],} 
$$
is $Y_{nm}(\cdot)$ \cite{williams2000fourier} of order $n$ and mode $m$
evaluated at $P$ directions,
$$
{\tiny
\mathbf{a}=
\left[\begin{array}{ccc}
\alpha_{00}(1) & \cdots & \alpha_{0 m}(L)\\
\vdots & \ddots & \vdots \\
\alpha_{n m}(1) & \cdots & \alpha_{n m}(L)
\end{array}\right]},
$$
is a ${(N_H+1)^2\times L}$ matrix containing the SH coefficients.
To obtain accurate coefficients, the number of HRTF measurements should greatly exceed the SH decomposition order, $P \gg\left(N_H+1\right)^2$.  $\mathbf{a}$ can be calculated through
\begin{equation}
\mathbf{a}=\left(\mathbf{Y}^{\intercal} \mathbf{Y}\right)^{-1} \mathbf{Y}^{\intercal} \mathbf{H} .
\label{eq:SHT}
\end{equation}

\begin{figure*}[t]
\begin{center}
\includegraphics[width=16cm]{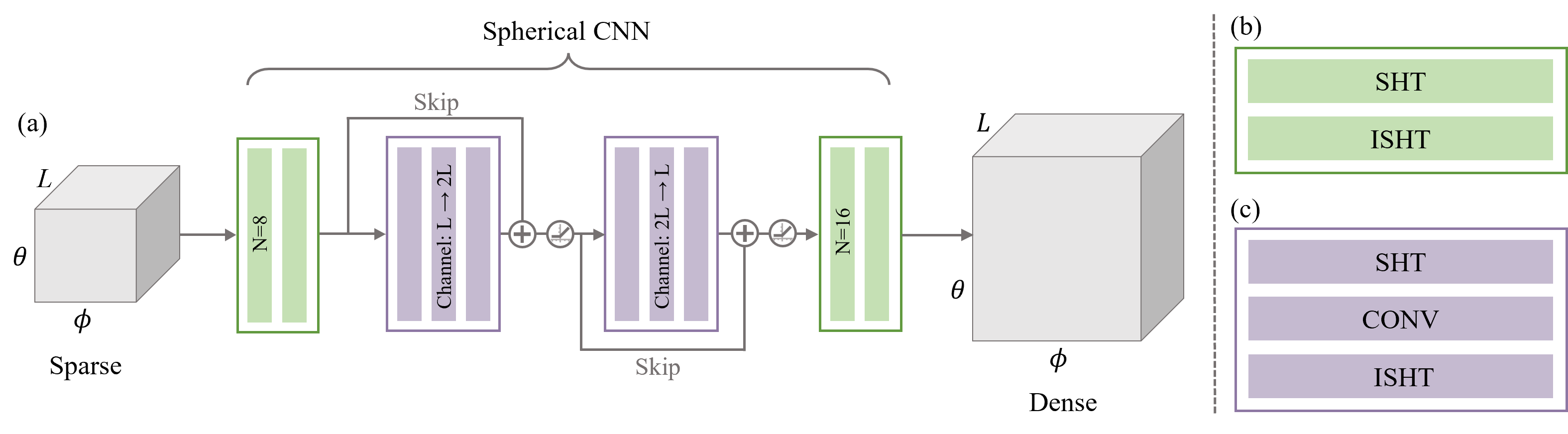}
\end{center}
\vspace{-0.5cm}
\caption{The network structures of the model. 
(a): It starts with a Mapping Block for sparse to dense signal transformation, followed by two Convolutional Blocks for spatial feature learning, and ends with a Mapping Block for dense domain mapping.
(b): The Mapping block: processes the spherical signal through SHT to convert it into SH coefficients, followed by ISHT maps back to the spherical signal region.
(c): The Convolutional block: performs SHT on the spherical signal, followed by coefficients multiplication and ISHT.}
\label{fig:modle}
\end{figure*}

Secondly,  we multiply the HRTF SH coefficients  $\mathbf{a}$ 
with a convolution kernel $k$ by Hadamard product, $\mathbf{a}\odot\mathbf{b}$, 
where $\mathbf{b}$ is 
$$
{\tiny
\mathbf{b}=
\left[\begin{array}{ccc}
\beta_{00}(1) & \cdots & \beta_{0 0}(L)\\
\vdots & \ddots & \vdots \\
\beta_{n 0}(1) & \cdots & \beta_{n 0}(L)
\end{array}\right]}
$$
where $\beta_{n 0}$ is the $m=0$ coefficient of kernel $k$. We realize  Eq.~\eqref{eq:conv} by a linear combination of spherical harmonics and associated  coefficients $\alpha_{n m}$ and $\beta_{n 0}$ \cite{driscoll1994computing},
\begin{equation}
(f * k)(\Omega)=\sum_{n=0}^{\infty} \sum_{m=-n}^n 2 \pi \sqrt{\frac{4 \pi}{2n+1}} \alpha_{n m} \beta_{n 0} Y_{n m}(\Omega) .
\label{eq:conv2}
\end{equation}

Eq.~\eqref{eq:conv2} represents the convolution of a single kernel whose output is summed across the rows $\sum_{i=1}^{L}(\mathbf{a}\odot\mathbf{b})_i$ to a ${(N_H+1)^2}\times1$ column vector. In a convolutional layer, there can be $u$ kernels and the layer output is $u$ column vectors concatenated to form $u$ channels
\begin{equation}
{\tiny
\mathbf{c} = [(\sum_{i=1}^{L}(\mathbf{a}\odot\mathbf{b})_i)_1, (\sum_{i=1}^{L}(\mathbf{a}\odot\mathbf{b})_i)_2, 
 \ldots, (\sum_{i=1}^{L}(\mathbf{a}\odot\mathbf{b})_i)_u] .}
\label{eq:nfeature}
\end{equation}

Thirdly, we use ISHT to convert the multiplication back into the spherical domain
\begin{equation}
\mathbf{Y} \mathbf{c}.
\label{eq:ISHT}
\end{equation}


\subsection{Model}

\noindent
\textbf{Network structure:}
The model structure is presented in  Fig. \ref{fig:modle} (a).
Overall this model generates the output block containing the densely interpolated HRTF 
from an input block containing the sparsely measured HRTF through a learnable kernel.  

The input and output are connected by the mapping block Fig.~\ref{fig:modle} (b)
and the convolution block Fig.~\ref{fig:modle} (c).
The mapping block, Fig.~\ref{fig:modle}(c), comprises of the SHT and the ISHT, 
which  aggregates the learned feature and maps back to the $\left\{H_{\text{M}}(\Omega, l)\right\}_{\Omega \in \Omega^{\text{dense}} }$ domain.
The convolutional blocks,  Fig. \ref{fig:modle}(c), 
consist of SHT (Eq. \ref{eq:SHT}), a convolutional layer (Eq. \ref{eq:nfeature}), and ISHT (Eq. \ref{eq:ISHT}). The sparse HRTFs undergo two convolutional blocks, which contain skip connections \cite{he2016deep} and rectified linear unit (ReLU) activation function \cite{nair2010rectified}. 


\noindent
\textbf{Loss function:}
The logarithmic spectral distortion (LSD) between the true $H_{\text{M}}$ and predicted $\hat{H}_{\text {M}}$ is used as the loss function of the model and performance measure across methods in Sec. \ref{sec:simulation},
\begin{equation}
\text {LSD}=\sqrt{\frac{1}{PL} \sum_{p=1}^P\sum_{l=1}^L\left|H_{\mathrm{M}}\left(\Omega_p, l\right)-\hat{H}_{\text{M}}\left(\Omega_p, l\right)\right|^2}.
\label{eq:LSD}
\end{equation}

\section{simulation}
\label{sec:simulation}
In this section, we conducted experiments on HRTF interpolation to assess the effectiveness of our proposed method, and compare it with the SH method \cite{ahrens2012hrtf}, the SH+DNN method \cite{xi2021magnitude}, HRTF field method \cite{zhang2023hrtf}, and the CNN+GAN method \cite{hogg2023hrtf}.

\subsection{Data pre-processing}
Considering that multiple data sets need to be consistent \cite{wen2023mitigating} and sub-models may appear \cite{pauwels2023relevance}, we chose to conduct experiments exclusively on the HUTUBS dataset due to its extensive subject coverage \cite{brinkmann2019hutubs}.
We employed 94 subjects with duplicates excluded \cite{brinkmann2019hutubs}. 
We used 77, 10, and 7 subjects to generate training, validation, and test data, respectively. 
We used ground truth 16 order SH coefficients to generate HRTFs from 480 directions and across 
$L = 93$ frequency bins, spanning from $172$ Hz to $16$ kHz. 


\subsection{Training}
Fig. 2 shows the direction $(\theta,\phi)$ of $120$ known HRTFs and $360$ unknown HRTFs.
The LSD (Eq.~\eqref{eq:LSD}) serves as the loss function for the training process. 
The training used the Adam optimizer \cite{adam} with a batch size of $14$. 
Our model was trained on an NVIDIA V100 GPU for about $700$ epochs, 
with early stopping determined by validation data.

\begin{figure}[t]
\begin{center}
\includegraphics[width=9cm]{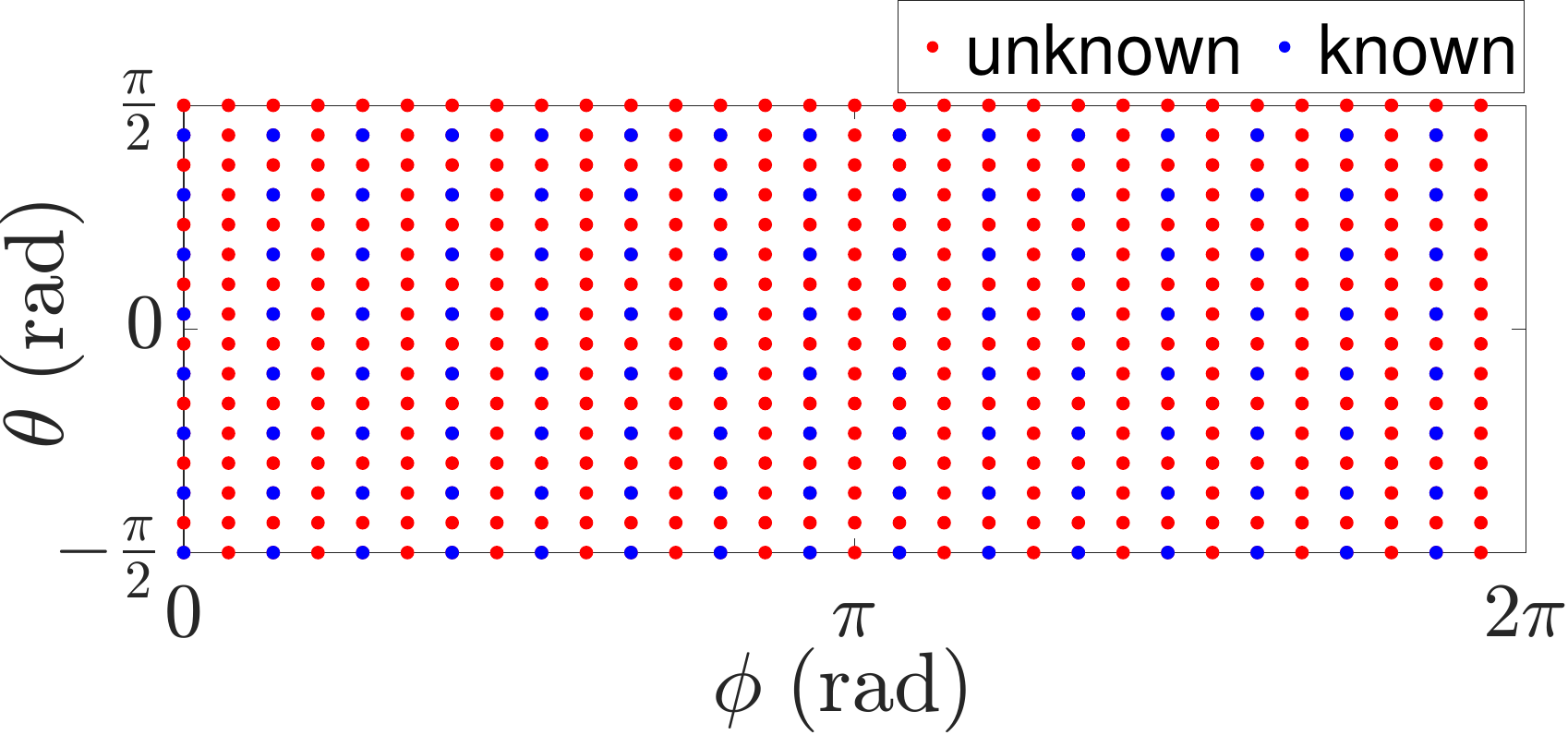}
\end{center}
\vspace{-0.3cm}
\caption{ Directions of 360 unknown and 120 known HRTFs.}
\label{fig:fig2}
\end{figure}

\begin{table}[t]
\centering
\caption{Comparison of model Parameters and average LSD for Unknown HRTFs across Subjects ID 88-94.}
\begin{tabular}{lcc}
\hline HRTF model & Parameters & Unknown \\
\hline SH $\text{N}=8$ \cite{ahrens2012hrtf} & - & 7.24 \\
SH+DNN \cite{xi2021magnitude} & 803241 & 5.11 \\
HRTF field \cite{zhang2023hrtf} & 4458589 & 4.19 \\
CNN+GAN \cite{hogg2023hrtf} & 100827332 & 4.36 \\
Proposed & \textbf{484530} & \textbf{2.17} \\
\hline
\end{tabular}
\end{table}

\begin{figure}[t]
\begin{center}
\includegraphics[width=9.0cm]{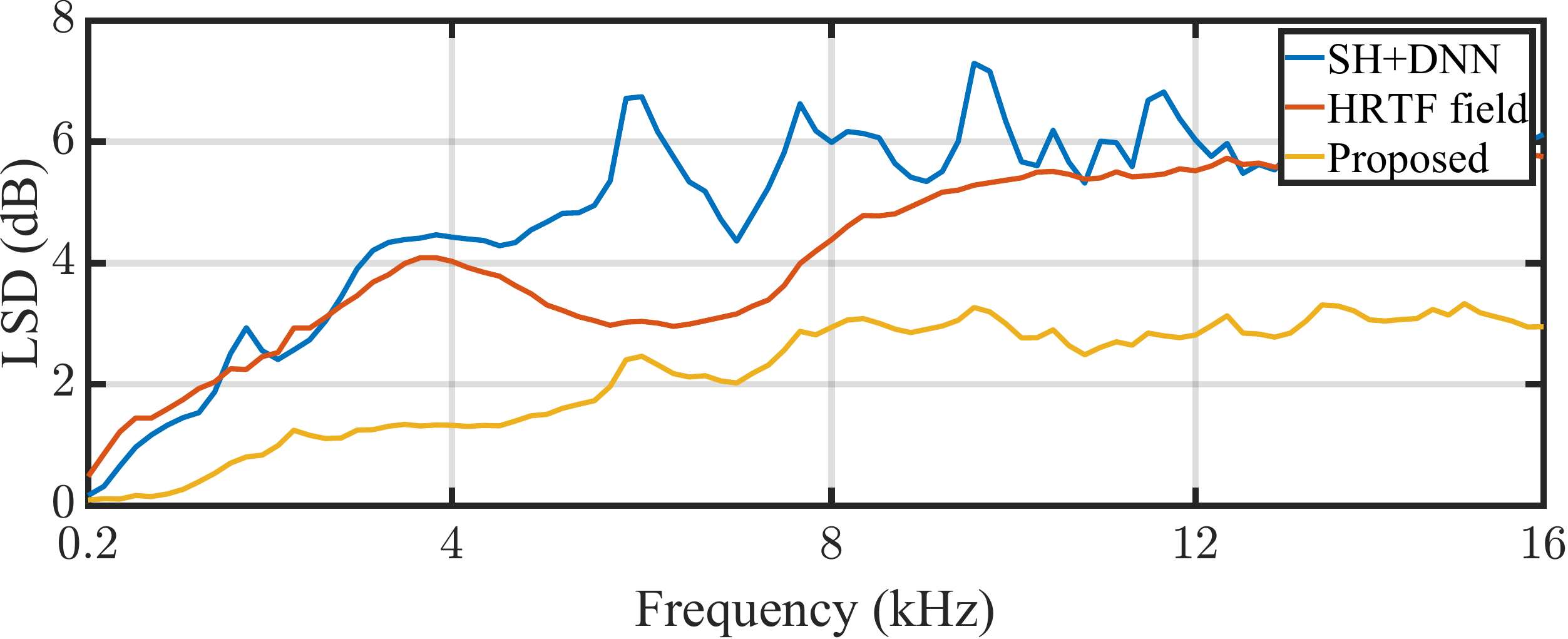}
\end{center}
\vspace{-0.4cm}
\caption{Interpolation LSD over frequencies.}
\end{figure}

\begin{figure}[t]

\begin{minipage}[t]{0.49\linewidth}
  \centering
  \centerline{\includegraphics[width=4cm]{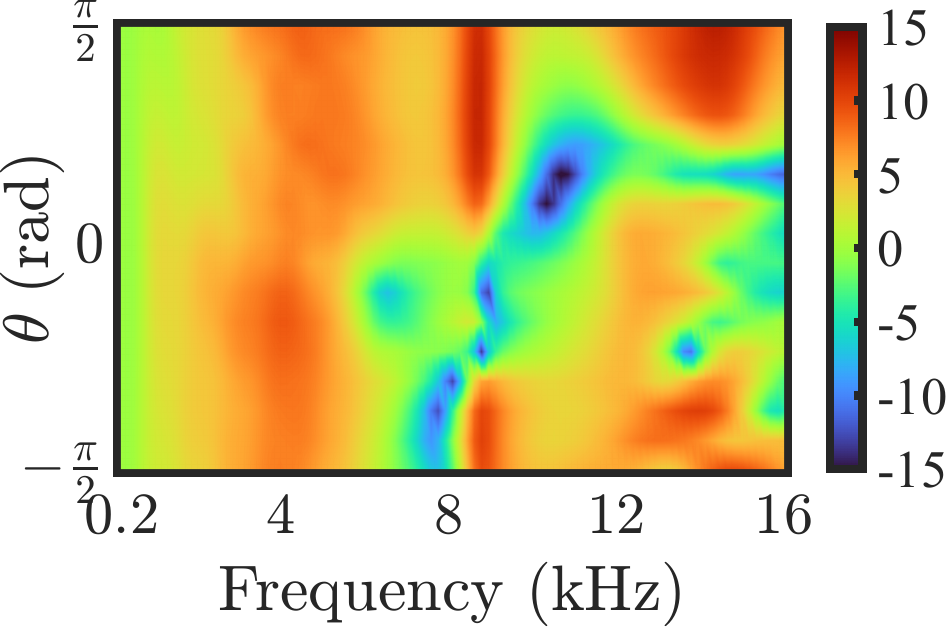}}
  \centerline{(a) Ground truth}\medskip
\end{minipage}
\begin{minipage}[t]{0.49\linewidth}
  \centering
  \centerline{\includegraphics[width=4cm]{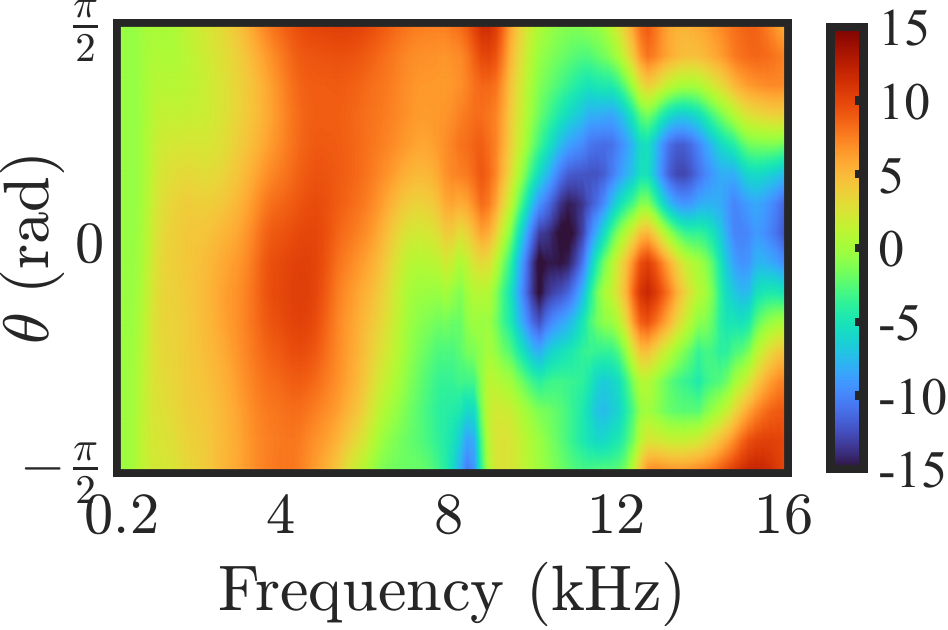}}
  \centerline{(b) SH+DNN} \medskip
\end{minipage}
%
\begin{minipage}[t]{0.49\linewidth}
  \centering
  \centerline{\includegraphics[width=4cm]{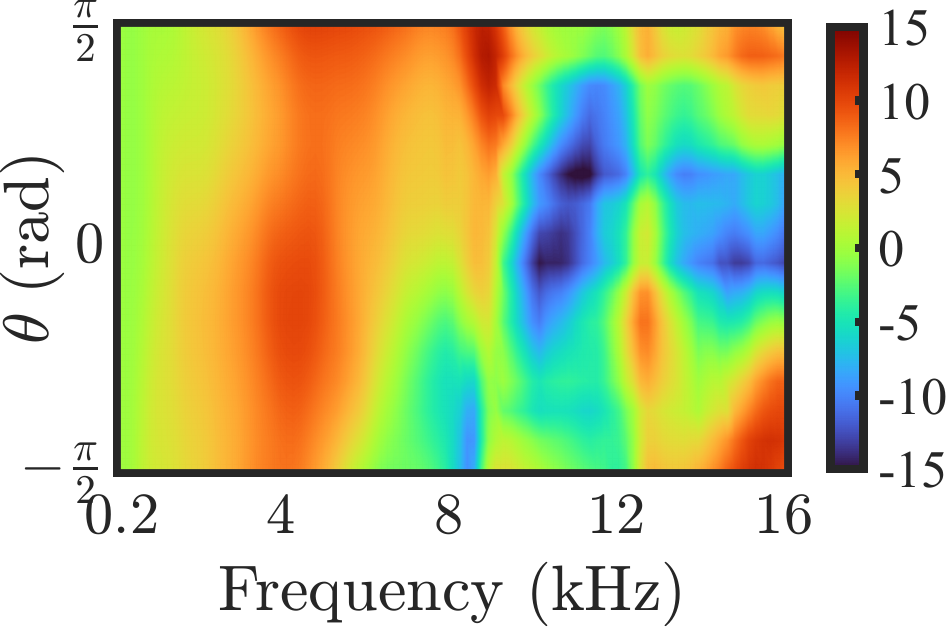}}
  \centerline{(c) HRTF field}\medskip
\end{minipage}
\begin{minipage}[t]{0.49\linewidth}
  \centering
  \centerline{\includegraphics[width=4cm]{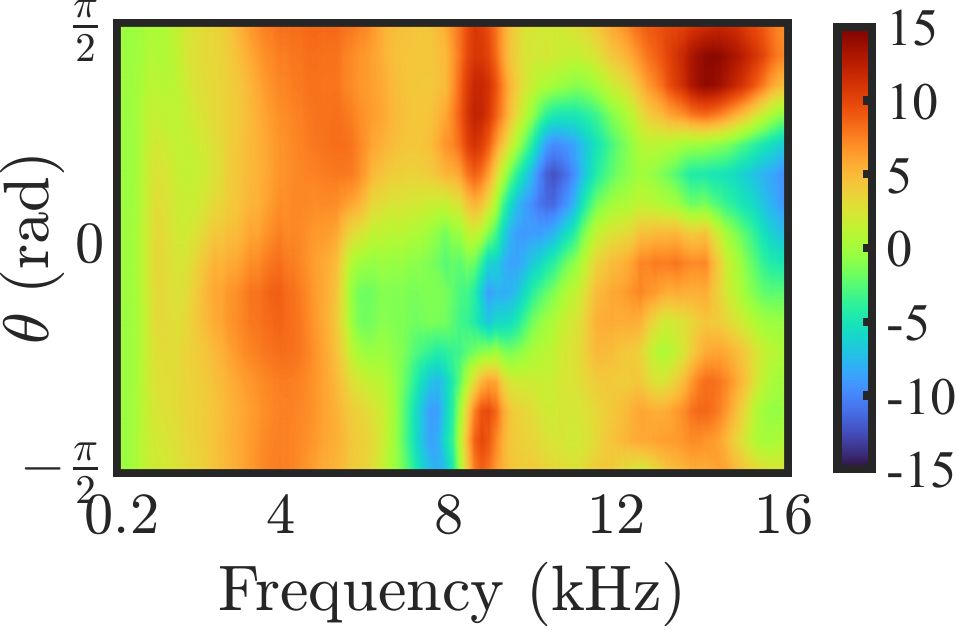}}
  \centerline{(d) Proposed}\medskip
\end{minipage}
%
\vspace{-0.3cm}
\caption{The left ear HRTF at $\phi = \pi$ of Subject ID 89.}
\label{fig:res}
\end{figure}

\subsection{Result and discussion}

We conducted a comparative analysis by adopting interpolation methods, including SH, SH+DNN, HRTF field, and CNN+GAN.
Table 1 provides insights into the number of parameters and average LSD of learning-based methods for unknown HRTFs across Subjects ID 88-94. 
Notably, the proposed method exhibits the lowest parameter count due to its streamlined architecture consisting of only two convolution blocks.
In contrast, both SH+DNN and HRTF field methods rely on fully connected networks. 
SH+DNN requires sub-networks training for each frequency bin while
CNN+GAN employs a generator with $11$ conventional CNN layers. 
This observation highlights the efficiency of our approach.
Furthermore, our method achieves the lowest LSD. 
The SH method with $8$th-order coefficients performed worse than SH+DNN with $4$th-order coefficients,
due to its inability to accurately compute $8$th-order coefficients with $120$ known HRTFs.
It indicates that the proposed method learns from convolution blocks, rather than solely relying on direct inferences through the mapping blocks.

We compare the HRTF interpolation performance of the proposed model with the SH+DNN and HRTF field model which has fewer parameters.
Interpolating the high-frequency part of HRTF (above $8$kHz) is challenging. Fig. 3 confirms that our method performs well in high frequencies.
Fig. 4 showcases HRTF interpolation at $\phi = \pi$. 
It provides a visual assessment of how well the methods fit the \textit{notch}, which represents a notable decrease at high-frequency, crucial for sound source direction detection \cite{li2020measurement}. 
The ground truth exhibits a peak in the 8 to 12 kHz frequency range. 
The other two methods show a different spectral shape compared to the ground truth. 
In contrast, our proposed method appropriately captures the peaks and notches of the ground truth.

\section{conclusion}
This paper proposed a spherical CNN method to interpolate HRTFs based on sparse measurements. 
By leveraging the SH transformation, the proposed method directly exploits spherical information, eliminating 
the need for plane-sphere projection. 
Simulation results demonstrated that our method outperforms both conventional SH methods and 
learning-based methods. 
This method not only enhances HRTF interpolation but also offers a compelling alternative 
to the widely employed SH method in soundfield reproduction tasks.

\clearpage
\vfill\pagebreak

\balance
\bibliographystyle{IEEEbib}
\bibliography{refs}

\end{document}